\documentclass[%
 reprint,
 amsmath,amssymb,
 aps,
]{revtex4-2}

\usepackage{graphicx}
\usepackage{dcolumn}
\usepackage{bm}

\begin{document}

\preprint{APS/123-QED}

\title{Photoassociation of multiple cold molecules in a dipole trap}
\thanks{A footnote to the article title}%

\author{Li Li}
\affiliation{CAS Key Laboratory of Quantum Information, University of Science and Technology of China, China}
\affiliation{CAS Center For Excellence in Quantum Information and Quantum Physics, University of Science and Technology of China, China}
\author{Yi-Jia Liu}
\affiliation{CAS Key Laboratory of Quantum Information, University of Science and Technology of China, China}
\affiliation{CAS Center For Excellence in Quantum Information and Quantum Physics, University of Science and Technology of China, China}
\author{Xiao-Long Zhou}
\affiliation{CAS Key Laboratory of Quantum Information, University of Science and Technology of China, China}
\affiliation{CAS Center For Excellence in Quantum Information and Quantum Physics, University of Science and Technology of China, China}
\author{Dong-Yu Huang}
\affiliation{CAS Key Laboratory of Quantum Information, University of Science and Technology of China, China}
\affiliation{CAS Center For Excellence in Quantum Information and Quantum Physics, University of Science and Technology of China, China}
\affiliation{Hefei National Laboratory, University of Science and Technology of China, China}
\author{Ze-Min Shen}
\affiliation{CAS Key Laboratory of Quantum Information, University of Science and Technology of China, China}
\affiliation{CAS Center For Excellence in Quantum Information and Quantum Physics, University of Science and Technology of China, China}
\author{Si-Jian He}
\affiliation{CAS Key Laboratory of Quantum Information, University of Science and Technology of China, China}
\affiliation{CAS Center For Excellence in Quantum Information and Quantum Physics, University of Science and Technology of China, China}
\author{Zhao-Di Liu}
\affiliation{CAS Key Laboratory of Quantum Information, University of Science and Technology of China, China}
\affiliation{CAS Center For Excellence in Quantum Information and Quantum Physics, University of Science and Technology of China, China}
\author{Jian Wang}
\email{jwang28@ustc.edu.cn}
\affiliation{CAS Key Laboratory of Quantum Information, University of Science and Technology of China, China}
\affiliation{CAS Center For Excellence in Quantum Information and Quantum Physics, University of Science and Technology of China, China}
\author{Chuan-Feng Li}
\email{cfli@ustc.edu.cn}
\affiliation{CAS Key Laboratory of Quantum Information, University of Science and Technology of China, China}
\affiliation{CAS Center For Excellence in Quantum Information and Quantum Physics, University of Science and Technology of China, China}
\affiliation{Hefei National Laboratory, University of Science and Technology of China, China}
\author{Guang-Can Guo}
\affiliation{CAS Key Laboratory of Quantum Information, University of Science and Technology of China, China}
\affiliation{CAS Center For Excellence in Quantum Information and Quantum Physics, University of Science and Technology of China, China}
\affiliation{Hefei National Laboratory, University of Science and Technology of China, China}

\date{\today}

\begin{abstract}
The generation of cold molecules is a core topic in the field of cold atoms and molecules, which has advanced relevant research like ultracold chemistry, quantum computation, and quantum metrology.
With high atomic phase space density, optical dipole trap has been widely performed to prepare and trap cold molecules, and can also be further developed for multiple cold molecule formation and dynamics study.
In this work, Rb$_2$ molecules are photoassociated in the magneto-optical trap to obtain precise rovibrational spectroscopy, which provides accurate numerical references for multiple photoassociations. 
By achieving the harsh requirements of photoassociation in the optical dipole trap, the cold molecule photoassociation process is well explored, and different rovibrational cold molecules are formed in the optical dipole trap for the first time.
This method can be universally extended to simultaneously photoassociate various molecules with different internal states or atomic species in just one optical dipole trap, and then advance generous cold molecule research such as cold molecule collision dynamics.  
\end{abstract}

\maketitle


\section{Introduction}
Cold molecules hold significant research potential and are widely considered for applications in ultracold chemistry, quantum optics, and quantum metrology~\cite{Hu,BJ,Sa}. 
With the low temperature of cold molecules, the energy resolution of chemical reaction processes has been greatly improved and explored~\cite{Ose,QUE}, and can attain quantum regime like quantum degenerate gas~\cite{WYN,De}. 
Manipulation of cold polar molecules holding a long-range tunable dipole interaction is appealing for fundamental physics and quantum applications such as quantum information processing~\cite{DEm,SOn,BAo}.
However, due to the complex rovibrational energy levels, cold molecules usually cannot be prepared as cold atoms which possess a cycling transition~\cite{DUl,KRe}.

The primary approach involves associating cold molecules with cold atoms by controlling atomic collision, which can directly achieve the low-temperature regime~\cite{CHa,BAl}.
The atom-molecule wavefunction can be either controlled by Feshbach resonance (magnetoassociation) or spin motion coupling, which forms weakly bound molecules in limited quantities~\cite{CHi,HE}. 
Noteworthy, by coupling atomic unbound states and molecular excited states with photons followed by spontaneous emission, photoassociation (PA) can continuously prepare more deeply bound even rovibrational ground states of cold molecules~\cite{DEI,KOc,BEL}.
This method has efficiently produced a variety of cold molecules and revealed PA spectroscopy that presents molecular energy level structures and calibrates related theoretical calculations~\cite{BER,JON,WUJ,ULM,GREE,RVA,PIC}.
The associated molecules typically spread across numerous vibrational levels and can be redistributed through optical pumping or coherently transferred to a single ground state by stimulated Raman adiabatic passage~\cite{VIT,AIK,BAR,LEU,YUY}.
Furthermore, extensive research has been conducted on the controlled collision of ultracold atoms and molecules using PA resonance, including the synthesis of triatomic molecules~\cite{PER,TAI,HOF,ELK}.

Optical dipole trap (ODT) is a powerful platform for cold atom and molecule experiments, including the preparation of quantum degenerate gas and the investigation of cold chemical reactions~\cite{GRE,Ger,PAR,LIL}.
With high atomic phase space density and long trapping lifetime, ODT is suitable for preparing and trapping cold molecules, followed by exploring complex cold molecule dynamics~\cite{MIL,MEN,CHE,WOL,FENG,LIU,PAS}. 
Especially, as shown in Fig.1, simultaneous photoassociation of various molecules with different internal states like long-range and short-range states of Rb$_2$~\cite{BEL,CHE} and Sr$_2$~\cite{LEU}, or different atomic species like LiCs~\cite{DEI}, YbLi~\cite{GREE}, NaLi~\cite{RVA}, KRb~\cite{AIK}, NaCs~\cite{YUY} and so on,  inside just one optical dipole trap is more valuable and attractive.
The produced cold molecules could be trapped in ODT, thus facilitating extensive research in the cold molecule and cold chemistry~\cite{Hu,BJ,Sa}, such as many-body cold molecule formation and collision dynamics~\cite{FENG,GAC,PEP}, and even cold chemistry reactions involving different molecular quantum degenerate gas~\cite{WYN,KOc,ULM}.
Nevertheless, photoassociation in ODT imposes stringent demands, which rely on the power density of PA laser and the overlap between PA lasers and cold atoms. 
Multiple photoassociations processes in ODT demand much higher challenges in meeting more complex requirements, thereby constraining the formation of various cold molecules inside just one optical dipole trap.

In this paper, the generation of different rovibrational cold molecules is first realized in ODT.
Firstly, the $0^+_u$ state of Rb$_2$ molecule is selected to acquire a much more precise PA spectroscopy than before, yielding accurate numerical references for the multiple photoassociations. 
Then the PA lasers of different wavelengths and ODT laser are combined using an endlessly single-mode fiber, and cold atoms are subsequently loaded into ODT for photoassociation that is characterized by atomic trap loss. 
By satisfying the requirements above, cold molecule photoassociation process in ODT is well explored.
All these results that different rovibrational cold molecules are formed in ODT for the first time.

\begin{figure}[tb]
	\centering
	\includegraphics[width=1.0\linewidth]{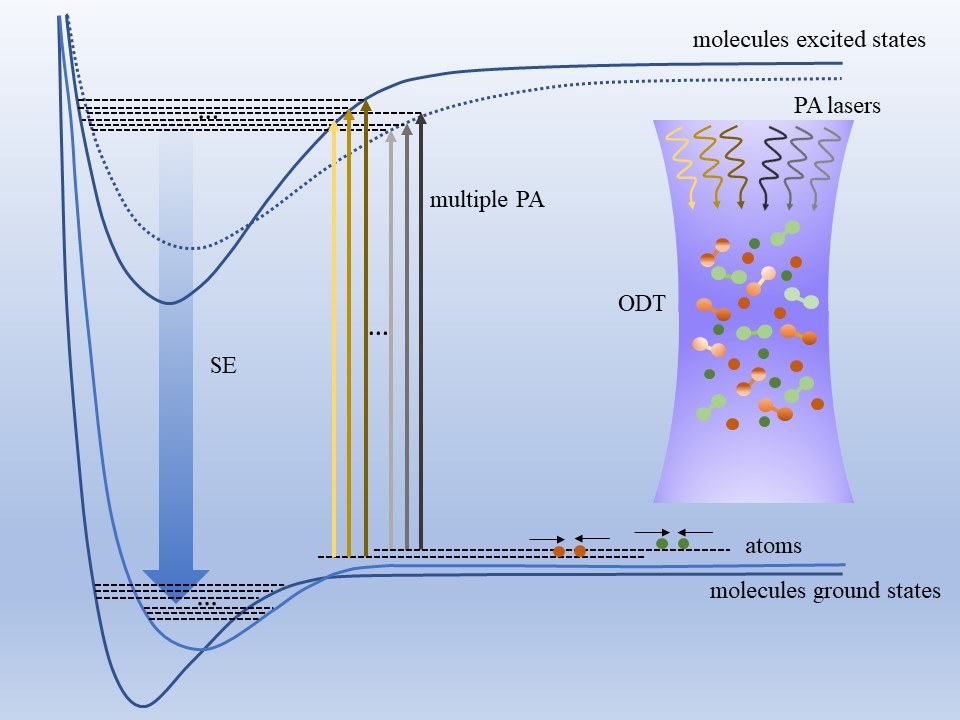}
	\caption{Schematic diagram of multiple PA processes. Various molecules with different internal states or atomic species are photoassociated in just one optical dipole trap and spontaneously radiate to corresponding ground states, and then are trapped in ODT, which can be used for research on cold molecules. SE: spontaneous emission.}
	\label{fig:1}
\end{figure}

\section{PA rovibrational spectroscopy}
Cold $^{85}$Rb atoms are prepared by the magneto-optical trap (MOT) in D2 line with a temperature of 150 $\mu$K. 
The trapping laser is 12 MHz red detuned from the cooling transition $|5S_{1/2}, F=3\rangle\rightarrow|5P_{3/2},F'=4\rangle$, and the repumping laser is locked to the transition $|5S_{1/2},F=2\rangle\rightarrow|5P_{3/2},F'=3\rangle$.
A 780 nm laser resonant with cooling transition is expanded to 2 mm and then used to image and characterize the cold atomic cloud by absorption imaging. 
Afterward, the PA laser is emitted by a continuous tunable Ti:sapphire laser (MBR 110, Coherent) with a power of 2 W, and coupled into the same fiber of the resonant laser to overlap with MOT. 
To measure the atomic trap loss caused by photoassociation, the atomic fluorescence is collected by an objective lens and focused on an avalanche photodiode.
Then a lock-in amplifier (MFLI-500 KHz, Zurich Instruments) is utilized to improve detection sensitivity by modulating the cooling laser at a frequency of 4 KHz, which could filter system noise and provide a sufficiently visible fluorescence signal as shown in Fig.2.

\begin{figure}
	\centering
	\includegraphics[scale=0.16]{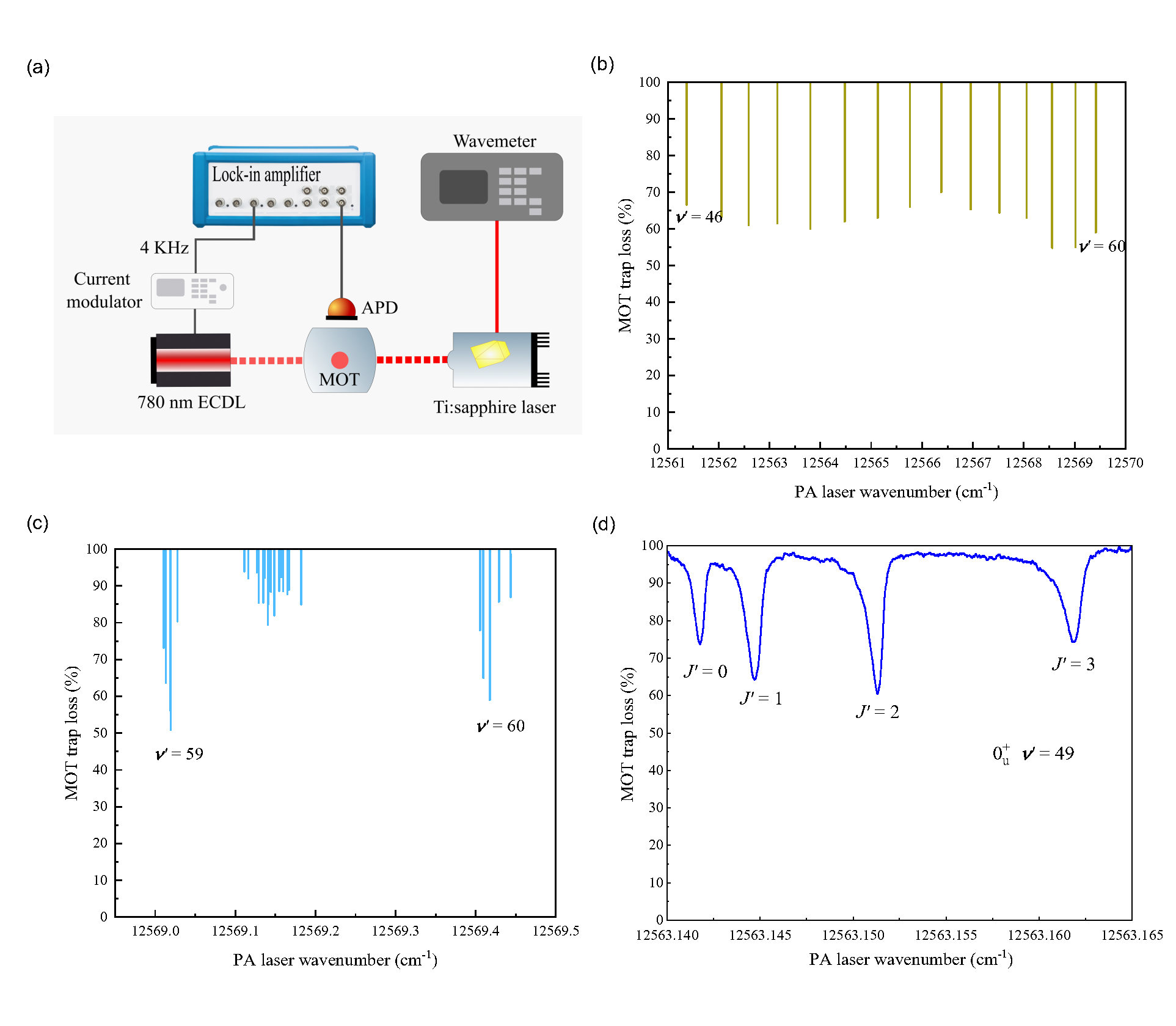}
	\caption{(a) Diagram of the scanning setup based on lock-in amplifier technology and a high-precision wavemeter. ECDL: external cavity diode laser, APD: avalanche photodiode. (b) Vibrational levels from $\nu'$ = 46 to $\nu'$ = 60 of $0^+_u$ state. The atomic fluorescence signal of MOT would decrease when the frequency of the Ti:sapphire laser resonates with the PA energy level. (c) Near the high vibrational levels, there are a lot of small dips of other molecular states. (d) Rotational levels of $\nu'$ = 49. The detailed data is presented in supplementary.}
	\label{fig:2}
\end{figure}

The PA laser is monitored by a high-precision wavemeter with an absolute accuracy of 30 MHz (WS7-30, HighFinesse), and scanned below the atomic 5S$_{1/2}$ + 5P$_{1/2}$ asymptotic limit to get the rovibrational spectroscopy of Rb$_2$ molecules in $0^+_u$ state. 
As an indication of molecule formation, the atomic fluorescence signal would decrease when the frequency of the Ti:sapphire laser resonates with the PA energy level.
In Fig.2, vibrational levels are identified where the trap losses are almost above 30$\%$, while higher atomic loss occurs due to the heating of near-resonant laser at $\nu'$ = 59 and $\nu'$ = 60.
Between the high vibrational levels, there are a number of small dips corresponding to other molecular states.
Besides, rotational levels of $\nu'$ = 49 express different trap losses, which are suitable for measuring the formation of different rovibrational cold molecules as below.
The much more precise spectroscopy than before~\cite{BER,CHE}, provides accurate numerical references for related theoretical calculations and experiments of Rb$_2$ molecules.

\begin{figure*}[t]
	\centering
	\includegraphics[width=0.75\linewidth]{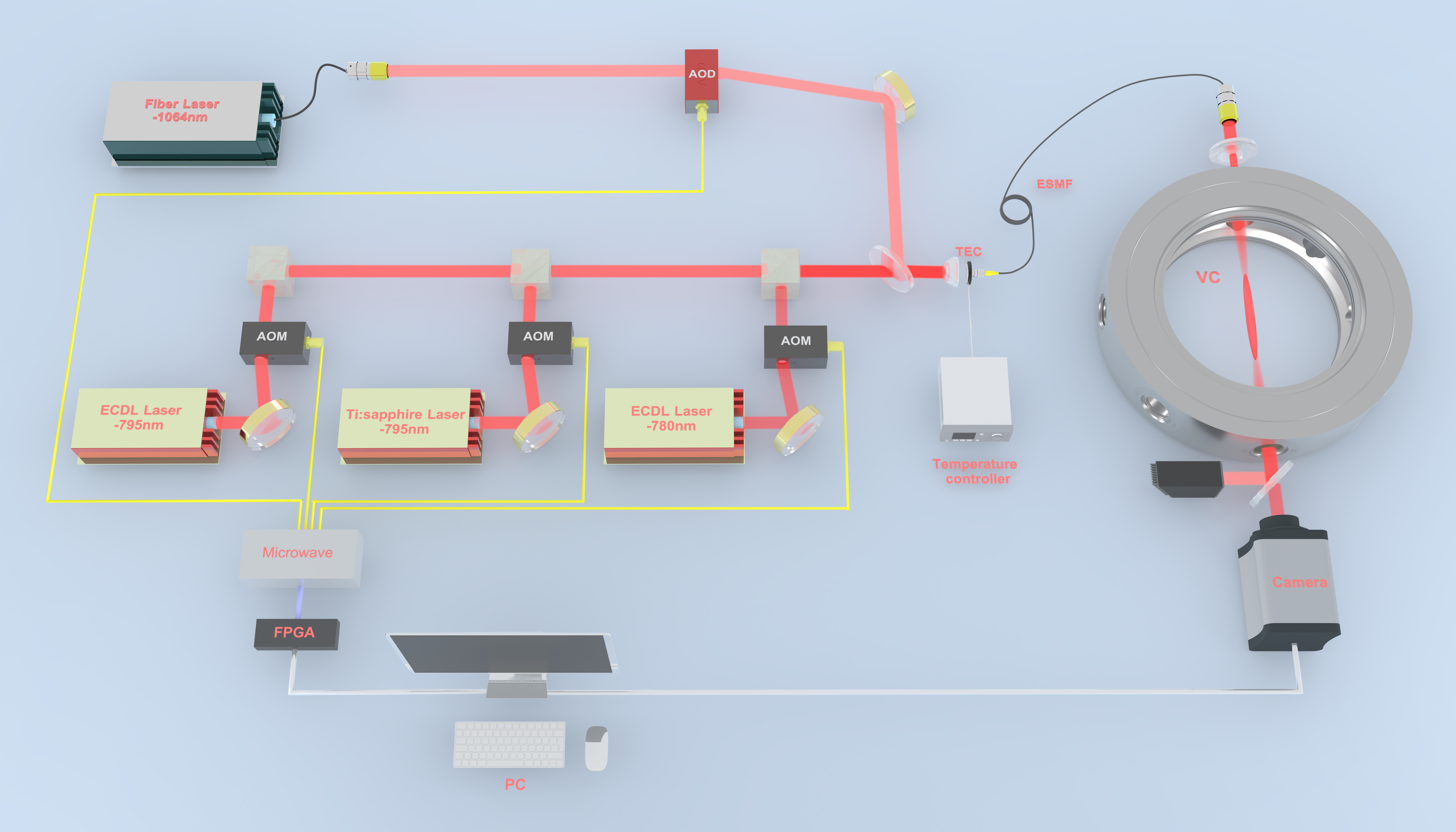}
	\caption{The experimental setup of multiple photoassociations process. The ODT laser at 1064 nm, PA lasers and 780 nm laser are coupled into the endlessly single-mode fiber and focused on MOT. ESMF: endlessly single-mode fiber, AOD: acousto-optic deflector, AOM: acoustic-optical modulator, TEC: thermoelectric cooler, VC: vacuum chamber, FPGA: Field Programmable Gate Array.}
	\label{fgr:3}
\end{figure*}

\section{Experimental setup}
Optical dipole trap is widely performed for cold atom and molecule experiments.
To reduce the atomic scattering and reach adequate potential well depth, a high-power fiber laser at 1064 nm is focused on MOT to load cold atoms, which is large detuning with the PA lasers.
The small beam waist and large power of the ODT laser induce high complexity and operational precision on photoassociation. 
Benefiting from the endlessly single-mode fiber (LMA-10, NKT-Photonics), the PA lasers of different wavelengths and ODT laser are coupled into the single fiber, which can maintain the Gaussian mode of large detuned lasers. 
Then cold atoms trapped in ODT are naturally completely covered by PA lasers after the fiber output as shown in Fig.3.

The ODT laser of 15 W is diffracted by an acousto-optic deflector to avoid AC stark shift. 
Besides, the high power would result in the heating of the fiber facet, thereby influencing the fiber coupling and subsequently affecting the stability of photoassociation.
Thus, an annular thermoelectric cooler and thermistor are affixed on the fiber adapter combined with a high-precision controller to maintain temperature. 
And fiber coupling efficiency of 75$\%$ is reached to reduce the heating of laser. 
Accordingly, the stability of the optical dipole trap has been greatly improved and photoassocaition can be well performed. 
In addition, the diameter of the PA laser beam is shaped to be suitable and coupled into the endlessly single-mode fiber with a similar coupling efficiency. 
The fiber output laser beams are expanded by a pair of achromatic lenses, and then focused on MOT to trap cold atoms and photoassociate cold molecules.
The whole experimental sequence requires fast and multi-channel operation which is implemented by FPGA.

\section{Formation of different rovibrational cold molecules in ODT}
Photoassociation rate in optical dipole trap can be approximatively expressed below,
\[R_{\mathrm{PA}} \propto \frac{I_{\mathrm{PA}} \cdot F \cdot n}{T^{1 / 2}},\]
where $I_{\mathrm{PA}}$ is the laser intensity, \emph{F} is the fixed Franck Condon factor denoted molecular transition strength and \emph{n} is the atomic density. 
Obviously, the rate would rise  as the temperature \emph{T} decreases, so polarization gradient cooling is performed to further cool the atoms after ODT loading as shown in Fig.4. 
The formation of different cold molecules is characterized by the trap loss of cold atoms, which are imaged by the 780 nm resonant laser. 
To obtain enough sensitivity, the resonant laser is along ODT to achieve sufficient atomic absorption in Fig.3.
The resonant laser is imaged by a high signal-to-noise ratio camera (ZYLA-5.5-CL10, Andor) that reads out the photon counts, and the absorption of cold atoms leads to a significant depression in the center of the spot in Fig.4.
Then the distribution of cold atoms in ODT is well characterized, which supports the exploration of the cold molecule photoassociation process.

\begin{figure}
	\centering
	\includegraphics[scale=0.15]{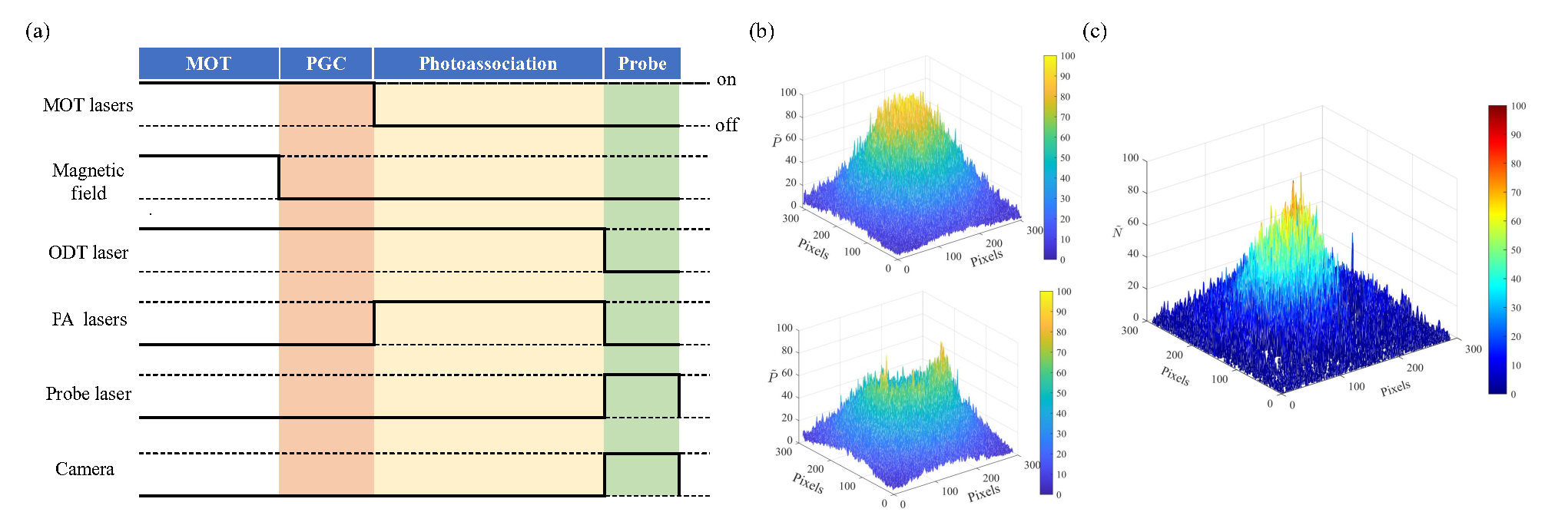}
	\caption{(a) The experimental time sequence from MOT preparation to probe with absorption imaging. PGC: polarization gradient cooling. (b) 780 nm resonant laser beam without or with cold atoms in ODT which induces a significant depression of power in the center of the laser spot. (c) The distribution of cold atoms that are trapped in ODT.}
	\label{fig:4}
\end{figure}

To simultaneously prepare different rovibrational cold molecules, the frequency of Ti:sapphire laser is resonant with the rotational energy level $J'$ = 2 of $\nu'$ = 49 to measure suitable conditions of the single photoassociation as shown in Fig.5.
As the time duration increases with a power of 50 mW, the trap loss also increases due to a higher probability of atomic collision, followed by a slight decrease that can be attributed to the limited trapping lifetime.
Following this, the power of the laser with a time duration of 500 ms is gradually increased, and the trap loss firstly goes up and then decreases due to the heating of the PA laser that is near-resonant with the D1 line.
Benefiting from the high atomic phase space density of ODT, the PA laser possesses a saturation power of only 40 mW after being focused.
An atomic trap loss of 30\% is realized handily with the naturally complete overlap, which is quite appropriate for preparing multiple cold molecules that require different PA lasers with large gaps in frequency.

Afterward, an external cavity diode laser at 795 nm is resonant with $J'$ = 1 of $\nu'$ = 49 which possesses a smaller association rate in Fig.2.
The time duration of photoassocaition is set to 500 ms, and the power of $J'$ = 2 is set at 22 mW while the other is scanned.
As shown in Fig.5, trap loss with a power of 30 mW results in a significant and stable increase compared to the photoassociation induced by individual lasers, indicating the production of cold molecules with different rotational energy levels.
As the power continues to increase, the process of photoassociation becomes constrained by atomic phase space density, and the jitter of photon counts from absorption imaging goes up due to the atomic heating caused by near-resonant PA laser.
Evidently, the association of cold molecules in $J'$ = 1 diminishes as a result of the generation of molecules in $J'$ = 2 due to cold atomic dissipation in ODT.
Then total number of molecules would be less than the sum of individual photoassociations at the same laser power in Fig.5.
The different cold molecules would spontaneously radiate to ground states and be trapped in ODT, which can be used for relevant cold chemical research.
Besides, by employing more advanced detection strategies, the preparation of more molecules with intricate internal states can be realized, facilitating extensive experimental studies of many-body molecular systems.

\begin{figure}
	\centering
	\includegraphics[scale=0.16]{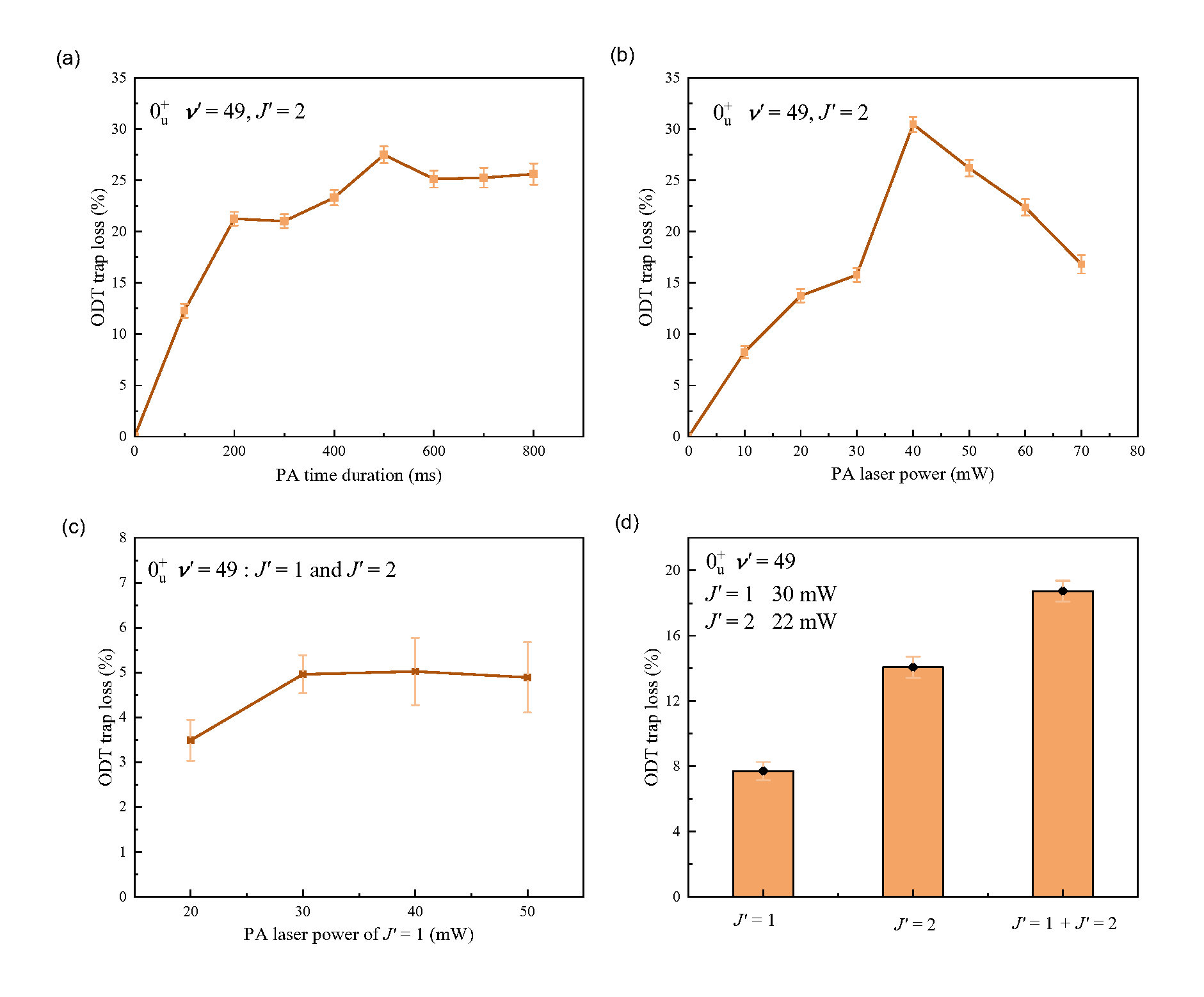}
	\caption{(a)(b) Trap loss of atoms in ODT changes with PA time duration and laser power. (c) Trap loss of the multiple photoassociations where the laser power of $J'$ = 1 is scanned to achieve a visible formation. (d) The laser power of $J'$ = 2 is 22 mW while the other is 30 mW presents a significant increase compared with each single photoassociation with the same laser power.}
	\label{fig:5}
\end{figure}

\section{Conclusions}
The much more precise spectroscopy of Rb$_2$ molecule in $0^+_u$ state is scanned out, which supports accurate numerical references for relevant chemical theoretical calculations and cold molecule formation.
After achieving the rigorous requirements of photoassociation in an optical dipole trap, the cold molecule photoassociation process is well explored, and then different rovibrational cold molecules are generated in ODT for the first time. 
Moreover, by guiding different PA lasers with large gaps in frequency, more cold molecules with different internal states or atomic species can be simultaneously generated inside just one optical dipole trap, followed by exploring generous cold molecule and cold chemistry research. 
This method could be universally extended to cold atom and molecule experiments based on ODT.

\section*{Acknowledgments}
The author thanks the Innovation Program for Quantum Science and Technology (No. 2021ZD0301200), the National Natural Science Foundation of China (Nos. 11804330 and 11821404), and the Fundamental Research Funds for the Central Universities (WK2470000038).

\end{document}